\documentclass[superscriptaddress,showpacs,preprintnumbers,amsmath,amssymb,twocolumn,prl]{revtex4}

\usepackage{graphicx}
\usepackage{dcolumn}
\usepackage{bm}
\usepackage{color}
\usepackage{braket}
\usepackage{ulem}

\newcommand{\comment}[1]{}
\def\be{\begin{equation}}
\def\ee{\end{equation}}
\def\bea{\begin{eqnarray}}
\def\eea{\end{eqnarray}}

\begin{document}

\title{The in-plane anisotropy of the hole $g$ factor in CdTe/(Cd,Mg)Te quantum wells\\ studied by spin-dependent photon echoes}

\author{S. V. Poltavtsev}
\email{sergei.poltavtcev@tu-dortmund.de}
\affiliation{Experimentelle Physik 2, Technische Universit\"at Dortmund, 44221 Dortmund, Germany}
\affiliation{Spin Optics Laboratory, St.~Petersburg State University, 198504 St.~Petersburg, Russia}
\author{I. A. Yugova}
\affiliation{Physics Faculty, St.~Petersburg State University, 199034 St.~Petersburg, Russia}
\author{A. N. Kosarev}
\affiliation{Experimentelle Physik 2, Technische Universit\"at Dortmund, 44221 Dortmund, Germany}
\affiliation{Ioffe Institute, Russian Academy of Sciences, 194021 St.~Petersburg, Russia}
\author{D. R. Yakovlev}
\affiliation{Experimentelle Physik 2, Technische Universit\"at Dortmund, 44221 Dortmund, Germany}
\affiliation{Ioffe Institute, Russian Academy of Sciences, 194021 St.~Petersburg, Russia}
\author{\\G. Karczewski}
\affiliation{Institute of Physics, Polish Academy of Sciences, PL-02668 Warsaw, Poland}
\author{S. Chusnutdinow}
\affiliation{Institute of Physics, Polish Academy of Sciences, PL-02668 Warsaw, Poland}
\author{T. Wojtowicz}
\affiliation{International Research Centre MagTop, Institute of Physics, Polish Academy of Sciences, PL-02668 Warsaw, Poland}
\author{I. A. Akimov}
\affiliation{Experimentelle Physik 2, Technische Universit\"at Dortmund, 44221 Dortmund, Germany}
\affiliation{Ioffe Institute, Russian Academy of Sciences, 194021 St.~Petersburg, Russia}
\author{M. Bayer}
\affiliation{Experimentelle Physik 2, Technische Universit\"at Dortmund, 44221 Dortmund, Germany}
\affiliation{Ioffe Institute, Russian Academy of Sciences, 194021 St.~Petersburg, Russia}

\date{\today}

\begin{abstract}
We use the two-pulse spin-dependent photon echo technique to study the in-plane hole spin anisotropy in a 20~nm-thick CdTe/Cd$_{0.76}$Mg$_{0.24}$Te single quantum well by exciting the donor-bound exciton resonance. We take advantage of the photon echo sensitivity to the relative phase of the electron and hole spin precession and study various interactions contributing to the hole in-plane spin properties. The main contribution is found to arise from the crystal cubic symmetry described by the Luttinger parameter $q=0.095$, which is substantially larger than the one theoretically expected for CdTe or found in other quantum well structures. Another contribution is induced by the strain within the quantum well. These two contributions manifest as different harmonics of the spin precession frequencies in the photon echo experiment, when strength and orientation of the Voigt magnetic field are varied. The magnitude of the effective in-plane hole $g$ factor is found to vary in the range $|\tilde{g_h}|$=0.125--0.160 in the well plane.
\end{abstract}



\maketitle

\section{I. Introduction}

Once the spin properties of a semiconductor quantum system become relevant, a detailed knowledge of the spin level structure of the confined carriers is essential. Particularly, for manipulating an optically excited state in a quantum well (QW) or in a quantum dot (QD), the properties of the hole spin must be taken into account. Due to the strong spin-orbit interaction, the structure of the valence band states is complex and, as a consequence, also the hole $g$ factor is strongly anisotropic \cite{SpinBook}. In magnetic field this is manifested in an anisotropic Zeeman splitting. Here, the configuration in which the magnetic field is applied perpendicular to the light wavevector (and to the structure quantization axis), for which the intrinsic optical transitions become linearly polarized, is especially interesting. In the isotropic case, the optical transitions split by the field would be polarized either along or perpendicular to the magnetic field axis as for atomic systems \cite{HakenWolfBook}. However, in a solid state, the effective magnetic field, acting on the holes, does not coincide with the external magnetic field. This is because confinement and strain significantly modify the valence band eigenstates and, as a  result, change the polarization of the optical transitions \cite{Semenov2003}.

There are various methods to study the hole spin in low-dimensional semiconductor systems. Some of them, such as polarized photoluminescence \cite{KusrayevPRL1999, KoudinovPRB2004, Bogucki2016} and spin-flip Raman scattering \cite{Sirenko1997, Kusrayev2002, Koudinov2006}, require the application of quite strong transverse magnetic fields in order to resolve the small Zeeman splittings, by which also band mixing effects are changed, for example. Additionally, tilting of the field axis away from the Voigt geometry is often used to involve the larger out-of-plane hole $g$ factor component in the Zeeman interaction \cite{Debus2013}. Also the spin noise technique allows studying the hole spin in certain systems with small or zero hole $g$ factor inhomogeneities such as single QDs \cite{Dahbashi2014}. Other methods based on pump-probe Kerr or Faraday rotation are mostly applicable to systems with resident holes, because, otherwise, the hole spin precession is screened by the signal originating from the electron spin \cite{Syperek2007, ZhukovPSS2014, Gradl2014}. As a result, the possibilities to study the in-plane anisotropy of the hole spin by these methods are somewhat limited and only a few of them have been employed so far for that \cite{Belykh2016, Kusrayev2002, Koudinov2006}.

On the other hand, time-resolved coherent optical methods allow for tracing the precise hole spin dynamics in the limit of small magnetic fields. One of these methods is four-wave mixing and especially photon echo technique, which provides access to the homogeneous optical linewidth \cite{AllenEberly}. Recently, it was shown for various systems that photon echoes depend sensitively on the applied transverse magnetic field, see e.g. \cite{Lisin2012, LangerPRL}. In particular, the in-plane hole $g$ factor can be measured using the spin-dependent photon echoes \cite{Invertor}. Therefore, exploiting coherent optical methods in combination with applying transverse magnetic fields opens unique opportunities to study the in-plane magnetic anisotropies of the electronic states in quantum confined semiconductor systems.

In this paper, we investigate the in-plane anisotropy of the hole spin in a single CdTe/(Cd,Mg)Te QW, manifested through the photon echoes from the neutral exciton bound to a donor (D$^0$X) in transverse magnetic field. Because of the absence of the electron-hole spin exchange interaction in this system, the observed state splittings increase linearly with the magnetic field strength. We find that the in-plane hole $g$ factor is highly anisotropic as a result of several contributions, which strongly influence the coherent optical dynamics of D$^0$X.

\section{II. Theoretical model}

We start with the theoretical consideration of the magnetic-field-dependent two-pulse photon echo (PE) generated by the isolated negatively charged exciton (trion) or D$^0$X in a QW, neglecting many-body effects. The system is excited by two short laser pulses of specific polarizations ($\mathcal{P}_1$ and $\mathcal{P}_2$) and separated by the time delay $\tau$. The excitation is followed by the two-pulse PE emission delayed by $\tau$ relative to the second pulse. The PE amplitude is studied as well in a specific $\mathcal{P}_{PE}$ polarization, so that the experiment can be characterized by the polarization sequence $\mathcal{P}_1 \mathcal{P}_2\rightarrow \mathcal{P}_{PE}$. We use as reference frame the one associated with the crystal axes, where $z$ $||$ [100] is the growth direction and $x$ $||$ [010], $y$ $||$ [001] are the in-plane axes, as shown in Fig.~\ref{axes}(a). The magnetic field is applied in the QW plane (${\bf B} \perp z$) under the angle $\varphi$ with respect to the $x$ axis.

\subsection{A. Circular polarization basis}

\begin{figure}[t]
	\includegraphics[width=\linewidth]{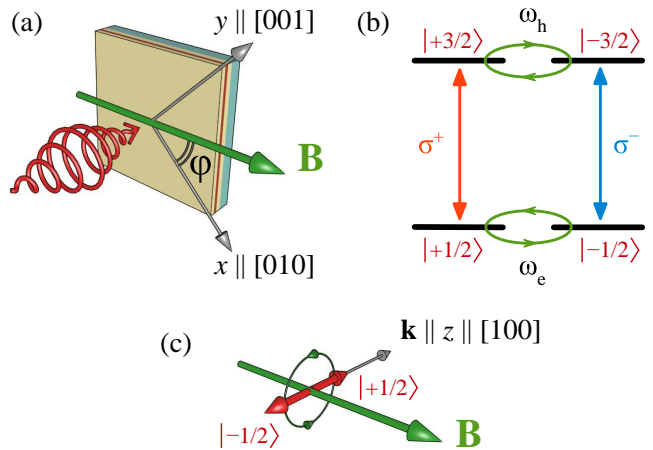}
	\caption{Representation of experimental geometry and spin level structure in the circular polarization basis: (a) Reference frame associated with the QW structure and its orientation with respect to the magnetic field. (b) Energy level scheme using as basis angular momentum states along ${\bf k} || z$. The numbers in brackets indicate the total angular momentum projections on the $z$ axis. (c) Electron spin orientations in the ground states $\ket{\pm1/2}$, which have equal energies and are mixed by the magnetic field.}
	\label{axes}
\end{figure}

The negatively charged exciton (trion) consists of two electrons and a hole, while the ground state implies a resident electron with spin 1/2. In the singlet state of D$^0$X, the two electron spins are antiparallel, so that the D$^0$X spin state is determined by the hole spin. Both systems, trion and donor-bound exciton, can be represented by the four-level energy scheme displayed in Fig.~\ref{axes}(b).

First, we consider the circular polarization basis, in which the eigenstates of total angular momentum projection on the $z$ axis correspond to the optically addressed states. The optical selection rules separate the four levels into two arms of allowed optical transitions using opposite circularly polarized light, as shown in Fig.~\ref{axes}(b). The electron ground states with angular momentum projections $\ket{\pm1/2}$ have the same energy and become mixed by the in-plane magnetic field ${\bf B}$. This is illustrated by the sketch in Fig.~\ref{axes}(c). So are the excited states (trion/D$^0$X) with total angular momentum projections $\ket{\pm3/2}$. 

To describe the magnetic properties of the electron or trion/D$^0$X states, we use the Hamiltonian in the form

\begin{align}
    \label{tensor}
	H = \frac{\omega}{2}
	\begin{pmatrix}
		0 & e^{-i\theta} \\
		e^{i\theta} & 0
	\end{pmatrix},
\end{align}

\noindent where $\omega$ is the effective Larmor frequency and $\theta$ is the angle between the $x$ axis and effective magnetic field axis, around which Larmor precession occurs. Because of the anisotropy of the spin splitting, the angle $\theta$ can differ from the actual magnetic field angle $\varphi$.

For the electron in the ground state, we take the Zeeman interaction to be isotropic using the constant in-plane electron Land\'e factor $g_e$. The Larmor frequency of the electron is $\omega_e=g_e\mu_B B / \hbar$, where $\mu_B$ is the Bohr magneton and $\hbar$ is the Planck constant. The electron experiences the actual magnetic field, so that $\theta = \varphi_e \equiv \varphi$.

The heavy hole spin is known to be strongly anisotropic in the QW plane \cite{KusrayevPRL1999}. Thereby, a number of interactions contributing to the effective hole $g$ factor $\tilde{g_h}$ have to be considered. We follow the theoretical approach by Semenov and Ryabchenko \cite{Semenov2003} and include the three main contributions expected to determine the hole spin dynamics: (i) the Zeeman interaction leading to a heavy hole splitting in third order of perturbation theory $\omega_z=\frac{3}{2}(g_h\mu_B B)^3/\hbar \Delta_{LH}^2$ \cite{BirPikusBook}, where $\Delta_{LH}$ is the heavy hole-light hole (HH-LH) splitting and $g_h$ characterizes the Zeeman interaction of the hole. The angle at which the effective magnetic field appears due to this contribution is $\theta=3\varphi$; (ii) the (non-Zeeman) interaction due to the cubic crystal symmetry with $\omega_q=\frac{3}{2}q\mu_B B/\hbar$ and $\theta=-\varphi$, where $q$ is the Luttinger parameter \cite{BirPikusBook, Marie1999}; (iii) the strain-induced potential of $C_{2v}$ symmetry for the hole inside the QW. This potential mixes HH and LH states in first order of perturbation theory and provides an additional splitting $\omega_{ht}=u\mu_B B/\hbar$ with the parameter $u\sim g_h/\Delta_{LH}$. This potential is determined by the in-plane strain axis with the angle $\phi$ relative to the $x$ axis. Summing these contributions results in the effective magnetic field angle $\theta=\varphi+2\phi+\pi/2$ \cite{Semenov2003}. Taking all contributions into account was shown to be important in self-assembled semiconductor QDs \cite{Kiessling2006, KoudinovPRB2004, Krizhanovskii2005}.

Comprising the three contributions, we write the hole spin Hamiltonian in the form

\begin{align}
\label{hole_H}
	H_h =& \frac{\omega_z}{2}
	\begin{pmatrix}
	0 & e^{-i3\varphi} \\
	e^{i3\varphi} & 0
	\end{pmatrix} + \frac{\omega_q}{2}
	\begin{pmatrix}
	0 & e^{i\varphi} \\
	e^{-i\varphi} & 0
	\end{pmatrix} +	\\
+ &\frac{i\omega_{ht}}{2}
	\begin{pmatrix}
	0 & -e^{-i(\varphi+2\phi)} \\
	e^{i(\varphi+2\phi)} & 0
	\end{pmatrix} \equiv \frac{\omega_h}{2}
	\begin{pmatrix}
	0 & e^{-i\varphi_h} \\
	e^{i\varphi_h} & 0
	\end{pmatrix}. \nonumber
\end{align}

The effective hole precession frequency $\omega_h=\tilde{g_h}\mu B/\hbar$ obtained from this is given by:

\begin{align}
	\label{omega_h}
	\omega_h = &\sqrt{N^2 + M^2},	\\
	N = \omega_z\cos(3\varphi) + \omega_q&\cos(\varphi) - \omega_{ht}\sin(\varphi + 2\phi), \nonumber \\
	M = \omega_z\sin(3\varphi) - \omega_q&\sin(\varphi) + \omega_{ht}\cos(\varphi + 2\phi).	\nonumber
\end{align}

\noindent The angle of the effective magnetic field experienced by the hole $\theta=\varphi_h$ can be found from $\tan(\varphi_h)=M/N$. 

\begin{figure}[t]
	\includegraphics[width=\linewidth]{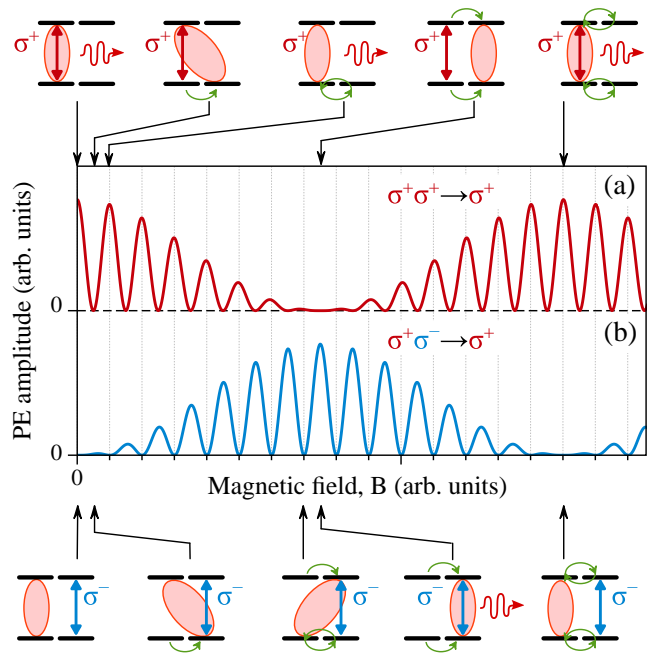}
	\caption{Magnetic field dependences of the PE amplitude calculated for the circularly (a) co-polarized and (b) cross-polarized configurations using Eq.~(\ref{circular_th}) for $|g_e|=15|\tilde{g_h}|$ and a fixed delay $\tau$. The cartoons above and below the plots indicate the involvement of the different states of the coherent ensemble in the four-level system in the PE formation at different magnetic field strengths. The red and blue arrows indicate the optical transitions addressed by excitation with the second pulse of certain circular polarization. The red ovals indicate the superposition state of the coherent ensemble. The PE, as illustrated by the wave arrow, can be emitted when the second pulse (polarization $\sigma^+$ (red) or $\sigma^-$ (blue)) can excite the coherent ensemble for which it has to be in the right superposition state.}
	\label{sigma_theor}
\end{figure}

As a result, we find for the PE amplitude analyzed in the polarization chosen to be identical to that of the first pulse ($\mathcal{P}_{PE}=\mathcal{P}_1$), for which the strongest and most informative PE signals are expected:

\begin{equation}
\label{circular_th}
P_{\sigma^+\sigma^\pm\rightarrow\sigma^+}\propto \big[1\pm\cos(\omega_e\tau)\big]\big[1\pm\cos(\omega_h\tau)\big].
\end{equation}

Hereafter, we fix the pulse delay $\tau$ and scan the magnetic field amplitude $B$, which allows us to neglect any relaxation processes in the coherent dynamics of the system. As a result, the effective Larmor frequencies $\omega_i$ are not constants in our consideration, but rather change with the magnetic field strength $B$ in accordance with the interactions (i)--(iii).

Figure~\ref{sigma_theor} shows the calculated magnetic field dependences of the PE amplitude in the circular polarizations basis. Initially, a coherent ensemble is created by the first $\sigma^+$ pulse, exciting a coherent superposition of the states $(\ket{+1/2},\ket{+3/2})$. Due to the electron spin precession, the low energy component of the superposition oscillates between the $\ket{+1/2}$ and $\ket{-1/2}$ ground states at the electron Larmor frequency $\omega_e$. Similarly, the high energy component oscillates between the $\ket{+3/2}$ and $\ket{-3/2}$ excited states at the effective hole Larmor frequency $\omega_h$. Various states of the coherent ensemble upon arrival of the second pulse are indicated by the red ovals in the cartoons of Fig.~\ref{sigma_theor}. Eventually, the PE is generated when the ensemble is in a state that can be excited by the second pulse: This is possible when either the $\sigma^+$ second pulse (red vertical arrow) hits the ensemble when it is in the $(\ket{+1/2},\ket{+3/2})$ superposition, or the $\sigma^-$ second pulse (blue vertical arrow) hits the ensemble in the $(\ket{-1/2},\ket{-3/2})$ superposition. As a result, the PE amplitude exhibits two types of oscillations when the field strength is varied. Typically $|g_e|>|\tilde{g_h}|$, therefore, the fast oscillations are due to the electron spin precession, while the slow oscillations of the envelope are due to the hole spin precession. We note that the signals carry only information about the absolute values of the effective electron and hole in-plane $g$ factors, $|g_e|$ and $|\tilde{g_h}|$, but are insensitive to the relative phase of the electron and hole spin precession.

\subsection{B. Linear polarization basis}

\begin{figure}[t]
	\includegraphics[width=\linewidth]{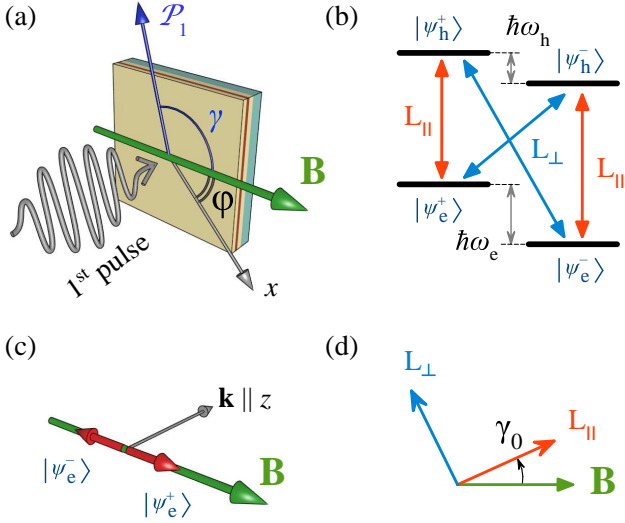}
	\caption{Representation of experimental geometry and spin level structure in the linear polarization basis: (a) Orientation of the first pulse polarization relative to the magnetic field $\bf B$. (b) Energy level scheme. (c) Electron spin orientations in the ground states $\ket{\pm1/2}$, which are split in energy and are not mixed by the magnetic field. (d) The eigenpolarizations $L_{||}$ and $L_\perp$ determined by the angle $\gamma_0$.}
	\label{axes_lin}
\end{figure}

In case of linearly polarized excitation, the first pulse polarization ($\mathcal{P}_1$) is characterized by the angle $\gamma$ relative to the magnetic field $\bf B$, as shown in Fig.~\ref{axes_lin}(a).

Here, to understand the coherent dynamics of the system, it is convenient to switch to the linear polarization basis shown in Fig.~\ref{axes_lin}(b). It involves as basis states $\ket{\psi^\pm_e}$ and $\ket{\psi^\pm_h}$ with spin projections parallel and anti-parallel to the effective magnetic field axis, respectively, and the two linear eigenpolarizations of the system, $L_{||}$ and $L_\perp$. The ground states $\ket{\psi^\pm_e}$ correspond to the two electron spin orientations along the $\bf B$ axis, as illustrated in Fig.~\ref{axes_lin}(c), with an energy splitting $\hbar\omega_e$. The excited states $\ket{\psi^\pm_h}$ with an energy splitting $\hbar\omega_h$ have spin orientations along the effective magnetic field that is directed under the angle $\varphi_h$ relative to the $x$-axis. The wavefunctions of these basis states are

\begin{align}
	\label{psi}
	\psi^\pm_e = \frac{1}{\sqrt{2}}\big(e^{-i\varphi_e/2}\ket{+1/2}&\pm e^{i\varphi_e/2}\ket{-1/2}\big), \nonumber \\
	\psi^\pm_h = \frac{1}{\sqrt{2}}\big(e^{-i\varphi_h/2}\ket{+3/2}&\pm e^{i\varphi_h/2}\ket{-3/2}\big)
\end{align}

\noindent with $\ket{\pm1/2}$ and $\ket{\pm3/2}$ being the circular basis states along the $z$ axis used before.

The orientation of the eigenpolarization $L_{||}$ is determined by the eigenangle $\gamma_0=(\varphi_h-\varphi_e)/2-\varphi + n\pi$, where $n$ is integer, as depicted in Fig.~\ref{axes_lin}(d).

\begin{figure}[t]
	\includegraphics[width=\linewidth]{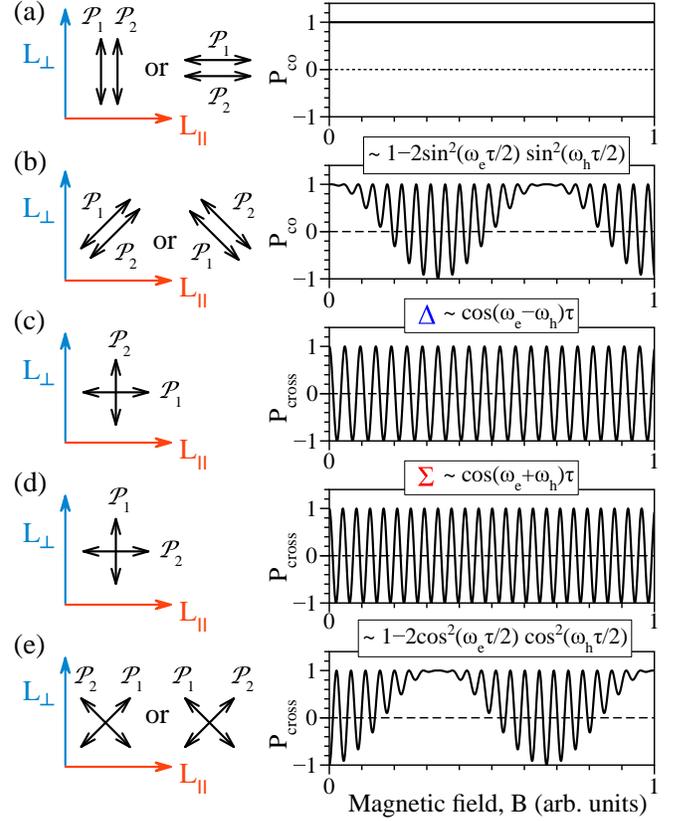}
	\caption{Magnetic field dependences of the PE amplitude calculated using Eqs.~(\ref{co_cross_th}), taking $g_e=15|\tilde{g_h}|$, for various linear polarization configurations in the eigenpolarization basis $L_{||}$,$L_\perp$ for a fixed delay $\tau$. (a)--(b) Co-polarized configurations. (c)--(e) Cross-polarized configurations. The symbols $\Delta$ and $\Sigma$ denote the PE signals oscillating at the difference and sum Larmor frequencies of electron and hole, respectively. The black arrows indicate the directions of linear polarization for the first and second pulses.}
	\label{linear_theor}
\end{figure}

Now we analyze the PE in the polarization of the first pulse ($\mathcal{P}_{PE}=\mathcal{P}_1$). The PE amplitudes evaluated for the co- and cross-polarized configurations are given by

\begin{align}
\label{co_cross_th}
P_{\text{co}}\propto 1 - 2\sin^2(&2\alpha)\sin^2(\omega_e\tau/2)\sin^2(\omega_h\tau/2), \\
P_{\text{cross}}\propto \cos\big[(\omega_e+&\omega_h)\tau\big]\sin^2(\alpha) +   \nonumber \\
+ \cos\big[(\omega_e-&\omega_h)\tau\big]\cos^2(\alpha) - \nonumber \\
- 2\sin^2(&2\alpha)\sin^2(\omega_e\tau/2)\sin^2(\omega_h\tau/2),  \nonumber 
\end{align}

\noindent where $\alpha = \gamma - \gamma_0$. We note that the PE detected in polarizations different from that of $\mathcal{P}_1$ contains higher harmonics so that it is harder to extract information from it.

\begin{figure*}[t]
	\includegraphics[width=\linewidth]{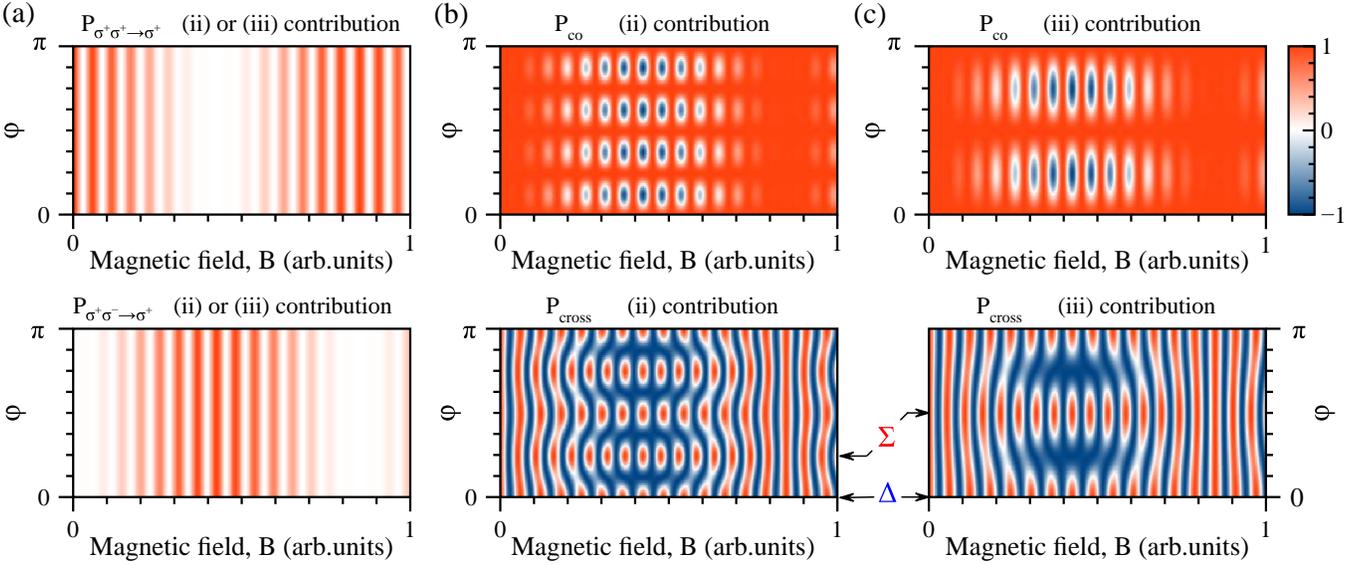}
	\caption{Spin-dependent PE as function of the magnetic field strength $B$ and orientation $\varphi$ calculated for different polarization configurations and contributions to the effective hole $g$ factor. Again, $g_e=15|\tilde{g_h}|$ is taken. (a) Contributions (ii) and (iii) in the circularly co-polarized (top) and cross-polarized (bottom) configurations. (b) Contribution (ii) in the linearly co-polarized (top) and cross-polarized (bottom) configurations. (c) Contribution (iii) in the linearly co-polarized (top) and cross-polarized (bottom) configurations. The symbols $\Delta$ and $\Sigma$ point at the PE signals oscillating at the difference and sum Larmor frequencies of electron and hole, respectively. The PE in the linear polarization configurations (b) and (c) is considered in the eigenpolarization basis, $\gamma = \gamma_0(\varphi=0)$.}
	\label{contributions}
\end{figure*}

Figure~\ref{linear_theor} summarizes the magnetic field dependences of the PE amplitude for the main co- and cross-polarized configurations with respect to the eigenpolarization basis $L_{||}$,$L_\perp$, as sketched in the left part of the figure. The PE amplitude is a constant in the co-polarized configurations $L_{||}L_{||}\rightarrow L_{||}$ and $L_\perp L_\perp\rightarrow L_\perp$, as shown in Fig.~\ref{linear_theor}(a). The PE amplitude oscillates as $\cos(\omega_e-\omega_h)\tau$ at the differential frequency ($\Delta$) and as $\cos(\omega_e+\omega_h)\tau$ at the sum frequency ($\Sigma$) in the cross-polarized configurations $L_{||}L_\perp\rightarrow L_{||}$ and $L_\perp L_{||}\rightarrow L_\perp$, respectively, as shown in Figs.~\ref{linear_theor}(c)--(d). Finally, Figs.~\ref{linear_theor}(b) and \ref{linear_theor}(e) display the magnetic field dependences when the pulses are diagonally co- and cross-polarized, respectively. Here, the fast oscillations originate from the electron spin precession, while the slow periodic amplitude modulation of these oscillations emerges from the hole spin precession. 

\subsection{C. Symmetry of different hole $g$ factor contributions}

Next we analyze how the various interactions contributing to the in-plane hole $g$ factor manifest in the spin-dependent PE signal when varying the magnetic field angle $\varphi$ in the QW plane. As before, we consider an isotropic electron in-plane $g$ factor, so that $\varphi_e=\varphi$.

When only a single interaction for the hole out of (i)--(iii) is present, the splitting of the heavy hole states will be isotropic, independent of the magnetic field angle $\varphi$. Therefore, the magnetic field dependences of the PE, measured in the $\sigma^+\sigma^\pm\rightarrow\sigma^+$ configurations, will exhibit no difference when $\varphi$ is varied, similar to the PEs shown in Fig.~\ref{sigma_theor}. This is illustrated in Fig.~\ref{contributions}(a) implying $\omega_h\propto B$. 

However, every of the considered interactions has a distinct symmetry which is reflected in the spin-dependent PE when measured in the linear polarization configurations. The Zeeman interaction (i), providing the $\omega_z\propto B^3$ splitting and entering the in-plane hole $g$ factor with the phase $\varphi_h=3\varphi$, results in $\alpha=\gamma$. Thus, the magnetic field dependences of the PE measured with linearly polarized pulses will be independent of the angle $\varphi$, and therefore the contribution (i) is isotropic. The contribution (ii) due to the cubic crystal symmetry with strength $\omega_q$ and phase $\varphi_h=-\varphi$ gives $\alpha=\gamma + 2\varphi$. As a result, the spin-dependent PE measured in the co-polarized configurations contains the eighth harmonic, while the PE measured in the cross-polarized configurations contains the fourth and eighth harmonics when varying the angle $\varphi$, as shown in Fig.~\ref{contributions}(b). Finally, the strain-induced interaction (iii) with strength $\omega_{ht}$, which enters the in-plane hole $g$ factor with phase $\varphi_h=\varphi+2\phi+\pi/2$, leads to $\alpha=\gamma+\varphi+\phi-\pi/4$. This contribution provides the fourth harmonics in the linearly co-polarized configurations, and the second and fourth harmonics in the linearly cross-polarized configurations as function of $\varphi$, as shown in Fig.~\ref{contributions}(c).

Superimposing these interactions in accordance with Eq.~(\ref{hole_H}) leads to interference effects, which in general makes the analysis of the experimentally measured spin-dependent PE a nontrivial task. When, however, one of the contributions (i)--(iii) prevails, the spin-dependent PE must exhibit certain symmetry properties, which help to identify the most important contribution and, as a result, simplify the analysis of the data.

\begin{figure*}[t]
	\includegraphics[width=\linewidth]{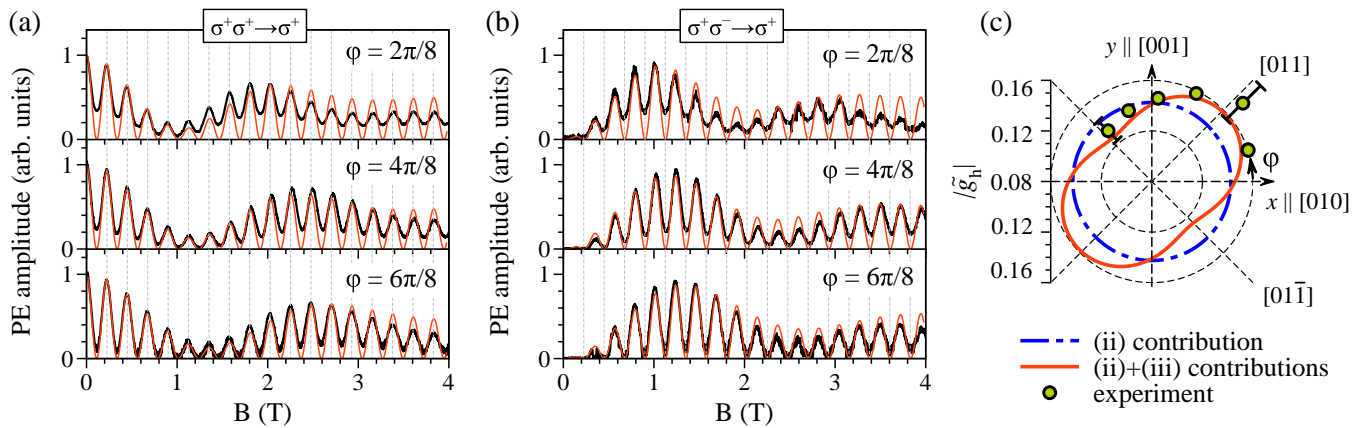}
	\caption{Summary of the experimental results on the spin-dependent PE measured on D$^0$X in a CdTe/Cd$_{0.76}$Mg$_{0.24}$Te QW at $T$=1.5~K for a pulse delay $\tau=200$~ps, using circularly polarized pulses: (a)--(b) Oscillations of the PE amplitude measured at $\varphi=2\pi/8, 4\pi/8,$ and $6\pi/8$ in the $\sigma^+\sigma^+\rightarrow\sigma^+$ and $\sigma^+\sigma^-\rightarrow\sigma^+$ polarization configurations, respectively. The thick black curves are the experimental data; the thin red curves are model calculations, with parameters adjusted to describe the complete data set. (c) Dependence of the absolute value of the effective in-plane hole $g$ factor $|\tilde{g_h}|$ on the in-plane angle $\varphi$ of the magnetic field orientation. The dots are results from the individual data fits; the red solid line gives the (ii) and (iii) contributions from the complete data set modeling; the blue dash-dotted circle is the (ii) contribution alone.}
	\label{data_sigma}
\end{figure*}

\section{III. Experimental results}

With the aim of studying the hole spin anisotropy by means of spin-dependent PE we used a 20~nm-thick CdTe/Cd$_{0.76}$Mg$_{0.24}$Te single QW (\#032112B). It was grown by molecular-beam epitaxy on a [100]-oriented GaAs substrate overgrown with 4.5~$\mu$m Cd$_{0.76}$Mg$_{0.24}$Te buffer and a short-period superlattice. The CdTe QW is sandwiched between 100~nm-thick Cd$_{0.76}$Mg$_{0.24}$Te barriers. The QW layer is unintentionally doped by donors ($n_d<10^{10}$~cm$^{-2}$) leading to the D$^0$X optical transition at the energy of 1.5973~eV ($T=1.5$~K), which we excite resonantly. The structure was examined before by various photon echo-based techniques and can be considered as model system \cite{SciRep2019, Poltavtsev2017, Salewski2017, Invertor}.

The experimental setup we employ here allows for time-resolved degenerate four-wave mixing (FWM) measurements. The sample was cooled down to the temperature of 1.5~K in a helium bath cryostat, equipped with a superconducting split-coil magnet. Two pulse trains from a Ti:sapphire laser with the spectral width of 0.9~meV (duration 2.3~ps) and wavevectors ${\bf k}_1$ and ${\bf k}_2$, oriented close to the sample normal, were focused into a spot of about $250$~$\mu$m on the sample. The second pulse train was delayed with an optical delay line by $\tau=200$~ps relative to the first one. The PE signal was collected in the reflection geometry along the $2{\bf k}_2-{\bf k}_1$ direction and mixed at the photodetector with a reference pulse, delayed by $2\tau=400$~ps with respect to the first pulse. The desired polarization of the detected FWM signal was chosen by the reference pulse polarization. The detected signal intensity is $I \propto \big|\text{Re}(E_{PE}E_{Ref}^*)\big|$, where $E_{PE}$ and $E_{Ref}$ are the electric field amplitudes of the PE and reference pulse, respectively. The energies per pulse were about 10~pJ, which corresponds to the pulse area of about $\pi/2$ or less \cite{Poltavtsev2017}. More details on the implemented technique can be found in Ref.~\cite{PSS2018}. The transverse magnetic field was applied in the sample plane at various angles $\varphi=n\pi/8$, $n$=1..6. The magnetic field strength was scanned in the range $B$=0--4~T.

Figures~\ref{data_sigma}(a)--\ref{data_sigma}(b) display PE measurements, performed in the $\sigma^+\sigma^+\rightarrow\sigma^+$ or $\sigma^+\sigma^-\rightarrow\sigma^+$ polarization configurations, for $\varphi=2\pi/8, 4\pi/8$, and $6\pi/8$. These data manifest two type of oscillations in full accord with our theoretical considerations, see Eq.~(\ref{circular_th}) and Fig.~\ref{sigma_theor}. The fast oscillations follow a frequency independent of the angle $\varphi$. From these oscillations the electron $g$ factor $|g_e|=1.584\pm0.005$ is extracted, in good agreement with previous studies \cite{Salewski2017,Saeed2018}. The envelope function of the fast oscillations exhibits slow oscillations caused by the hole spin precession. These slow oscillations appear to be in anti-phase for the $\sigma^+\sigma^+\rightarrow\sigma^+$ and $\sigma^+\sigma^-\rightarrow\sigma^+$ polarization configurations and demonstrate a $\varphi$-dependent period. From that period the effective in-plane hole $g$ factor $|\tilde{g_h}|=\hbar\omega_h/\mu_B B$ can be determined, as plotted with dots in Fig.~\ref{data_sigma}(c). It varies in the range $|\tilde{g_h}|$=0.13--0.17 with the highest value for the magnetic field oriented along the [011] crystalline axis.

\begin{figure*}[t]
	\includegraphics[width=\linewidth]{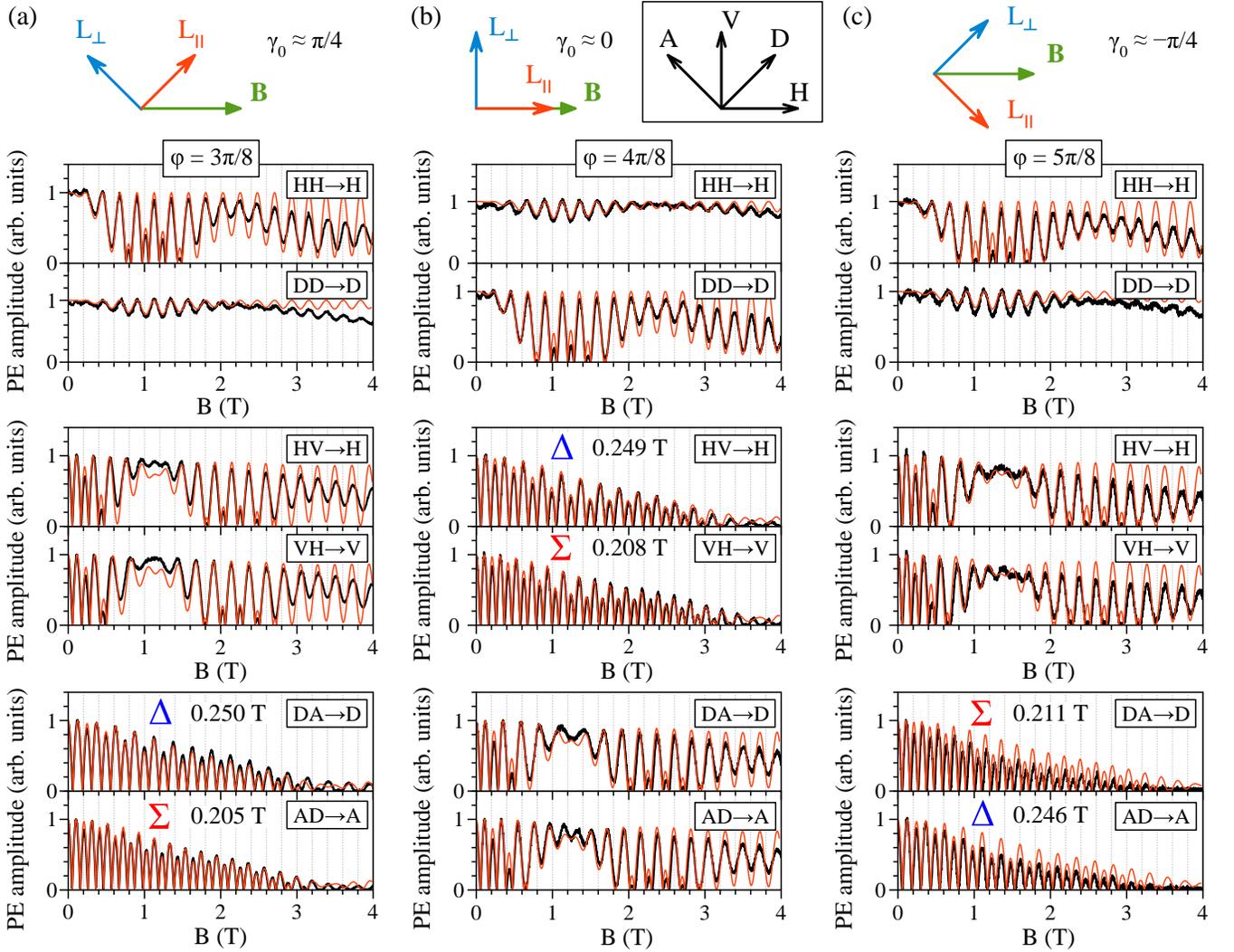}
	\caption{Spin-dependent PE measured with linearly polarized pulses at $\tau=200$~ps in various polarization configurations for different magnetic field orientation angles: (a) $\varphi=3\pi/8$; (b) $\varphi=4\pi/8$; (c) $\varphi=5\pi/8$. The experimental data (thick black curves) are normalized to unity. The thin red curves correspond to model calculations with parameters chosen to describe the complete data set. The schemes above the data depict the orientations of the eigenpolarizations $L_{||}$ and $L_\perp$. The arrows in the box give the linear polarization directions in the laboratory reference frame: H, D, V, and A correspond to the angles 0, $\pi/4$, $\pi/2$, and $3\pi/4$ with respect to the $\bf B$ axis. The numbers next to the $\Delta$ and $\Sigma$ symbols give the extracted difference and sum oscillation periods.}
	\label{data_linear}
\end{figure*}

Figure~\ref{data_linear} provides experimental data measured in the linear polarization configurations for the magnetic field orientations $\varphi=3\pi/8$, $4\pi/8$, and $5\pi/8$. In order to operate with specific linear polarizations in the laboratory reference frame associated with the magnetic field orientation we employ the following directions, as indicated in the inset of Fig.~\ref{data_linear}: Horizontal H$||\bf B$ ($\gamma=0$), vertical V$\perp$$\bf B$ ($\gamma=\pi/2$), diagonal D ($\gamma=\pi/4$), and anti-diagonal A$\perp$D ($\gamma=3\pi/4$). The data exhibit the following symmetry properties: In the co-polarized configurations HH$\rightarrow$H and DD$\rightarrow$D, the signal shape changes with the angle step $\varphi$ with an oscillation period of $\pi/4$, corresponding to the eighth harmonic. Similarly, in the cross-polarized configurations HV$\rightarrow$H (VH$\rightarrow$V) and DA$\rightarrow$D (AD$\rightarrow$A) periodic shape changes occur with the angle period of $\pi/2$, which corresponds to the forth harmonic. However, the anisotropic angular dependence of the effective hole $g$ factor [Fig.~\ref{data_sigma}(c)] evidences the presence of at least two different contributions, one of which is thus prevailing.

\section{IV. Analysis and discussion}

From the comparison of the data in Fig.~\ref{data_linear} with the magnetic field dependences of the PE in the eigen polarization basis (Fig.~\ref{linear_theor}), the orientation of this basis can be deduced: $\gamma_0\approx -2\varphi + n\pi$. We note that the measured signal is proportional to the absolute value of the PE amplitude, so that it should be compared with $|P_{\text{co}}|$ and $|P_{\text{cross}}|$ dependences. As a result, for the magnetic field tilted by $\varphi=3\pi/8$ the eigenangle is $\gamma_0\approx\pi/4$. Thus, the eigen polarizations $L_{||}$ and $L_\perp$ are aligned with the D and A polarizations, respectively, as shown schematically in Fig.~\ref{data_linear}(a). Similarly, the data measured at $\varphi=4\pi/8$ [Fig.~\ref{data_linear}(b)] have the eigenangle $\gamma_0\approx0$ and the $L_{||},L_\perp$ axes are aligned with the H and V polarizations. Accordingly, the data measured at $\varphi=5\pi/8$ [Fig.~\ref{data_linear}(c)] have the eigenangle $\gamma_0\approx-\pi/4$ and the $L_{||},L_\perp$ axes are aligned with the A and D polarizations.

We discuss now the details of the model adjusted to describe quantitatively all the collection of the experimental data. First, we neglect the Zeeman contribution (i) to the hole $g$ factor, since it is expected to be small ($\omega_z\sim\omega_h\times10^{-3}$) due to the sufficiently large HH-LH splitting of $\Delta_{LH}\approx15$~meV \cite{SciRep2019}. Moreover, it requires nonlinear dependence of the Zeeman splitting on magnetic field, which is hard to observe from our data. In order to take into account the inhomogeneity of the hole $g$ factor, we consider the Gaussian distribution of the two remaining contributions weights $q$ and $u$ with the same relative dispersions $\Delta p/p$ ($p=q,u$), for simplicity. We neglect the $g_e$ dispersion, which is expected to be within 1\% \cite{Saeed2018, Chamarro2019}.

From the theoretical modeling of the complete data set shown  with the red lines in Figs.~\ref{data_sigma}(a), \ref{data_sigma}(b) and \ref{data_linear} we find that the in-plane hole $g$ factor is dominated by the non-Zeeman contribution (ii) with the Luttinger parameter $q=0.095\pm0.005$. It is responsible for the main symmetry properties of the spin-dependent PE and corresponds to the effective hole $g$ factor $3/2q \approx 0.143$, shown in Fig.~\ref{data_sigma}(c) with the blue dash-dotted circle.

The strain-induced contribution (iii) has the weight of $u=0.016\pm0.006$ with the strain axis orientation $\phi=(94\pm2)^\circ$. This contribution interferes with the non-Zeeman contribution (ii) leading to the anisotropic angular dependence of the observed effective in-plane hole $g$ factor, shown in Fig.~\ref{data_sigma}(c) with the red solid line. The dependence is stretched along the [011] axis with the aspect ratio of about 30\% resulting in the hole effective $g$ factor varying in the range $|\tilde{g_h}|$=0.125--0.160.

The small amplitude oscillations observed in the linearly co-polarized configurations aligned with the $L_{||}$ axis, such as HH$\rightarrow$H at $\varphi=4\pi/8$ [Fig.~\ref{data_linear}(b)] or DD$\rightarrow$D at $\varphi=3\pi/8$ and $5\pi/8$ [Fig.~\ref{data_linear}(a), \ref{data_linear}(c)], are due to a deviation $\delta\varphi\sim4^\circ$ of the experimentally used sample orientations from the nominal $\varphi$ angles with respect to the actual crystal $x$ axis. This deviation was confirmed by the X-ray Laue analysis.

Finally, the dispersion of the hole $g$ factor can be characterized by the standard deviation $\Delta p/p\approx0.25$. It is responsible for the strong damping of the oscillating PE amplitude with the increase of the magnetic field strength $B$ in the $\Delta$ and $\Sigma$ polarization configurations such as HV$\rightarrow$H (VH$\rightarrow$V) at $\varphi=4\pi/8$ or DA$\rightarrow$D (AD$\rightarrow$A) at $\varphi=3\pi/8$ and $5\pi/8$ (Fig.~\ref{data_linear}). It causes also the contrast reduction in the PE amplitude oscillations in $\sigma^+\sigma^\pm\rightarrow\sigma^+$ polarization configurations (Fig.~\ref{data_sigma}).

There are other magnetic field-induced effects to mention, which we however neglected here. Because of the diamagnetic high energy shift of the D$^0$X spectral line ($\sim0.3$~meV at $B=4$~T) and subsequent detuning from the laser energy, the detected PE amplitude is somewhat reduced at $B>2$~T. Additionally, the strong magnetic field affects the spins dynamics during the optical pulse action, which may effectively reduce the detected PE amplitude.

As was mentioned before, the pulse delay $\tau$ can be varied at constant magnetic field in order to observe the spin-dependent PE. Thereby, the diamagnetic shift problem can be eliminated. However, since the PE amplitude experiences exponential decay with the optical coherence time ($T_2\approx100$~ps) \cite{Poltavtsev2017}, this has to be taken into account. The spin relaxation processes occur on the much longer times ($\geq1$~ns) and could be still neglected \cite{Salewski2017}.

The in-plane anisotropy of the hole $g$ factor studied by the magnetic-field-dependent photoluminescence in CdTe/(Cd,Mn)Te \cite{KusrayevPRL1999} and (Cd,Mn)Te/(Cd,Mn,Mg)Te QWs \cite{Koudinov2006} was described by the lowering to $C_{2v}$ crystal symmetry due to the strain-induced contribution (iii). In our study, however, we have shown that the cubic crystal symmetry plays dominant role in the anisotropic hole spin properties.

The magnitude of the Luttinger $q$ parameter in quantum wells was not experimentally studied well so far. It is usually disregarded in consideration because of its hypothetical smallness \cite{Lawaetz1971, Kiselev2001, Durnev2012}. In GaAs-based QWs, values in the range $q=0.01$--0.04 were evaluated from the photoluminescence studies \cite{Marie1999, Glasberg1999}. The $q$ parameter was not measured in CdTe/(Cd,Mg)Te QWs and, thus, it is difficult to conduct a comparison.

\section{V. Conclusions}

The photon echo-based technique is developed to monitor the precise spin-optical coherent dynamics of ensembles of the exciton complexes, which can be described by the four-level scheme. It may be applied to study charged excitons as well as neutral excitons bound to a donor or an acceptor in various semiconductor systems including quantum wells, quantum dot ensembles, and epilayers. The main advantage of the technique is the possibility to measure the in-plane components of the hole $g$ factor using relatively weak magnetic fields and to study various interactions of the hole spin with the in-plane magnetic field. In our study, the magnetic field as low as $B\approx200$~mT was sufficient to estimate the effective hole $g$ factor of about 0.1 from the spin-dependent photon echo measurements in certain polarization configurations at the pulse delay $\tau=200$~ps. This is because the effective hole $g$ factor can be evaluated from the deviation of the signal oscillation frequency from the electron Larmor frequency. Besides, the magnetic field strength can be scaled down further with the increasing $\tau$. Moreover, as compared with other methods studying the hole spin dynamics, it is not necessary to have the resident holes for that purpose. Eventually, applying the method to the CdTe/(Cd,Mg)Te quantum well we have been able to extract the parameter $q$ in the Luttinger-Kohn Hamiltonian, which is difficult to obtain by other optical techniques. Presented method can be extended for the magnetic fields tilted from the structure plane allowing thus for studying the out-of-plane hole spin properties.

\section{Acknowledgments}

The authors are thankful to Yu.G. Kusrayev and M.M. Glazov for fruitful discussions. The authors acknowledge the financial support by the Deutsche Forschungsgemeinschaft through the International Collaborative Research Centre TRR 160 (Project A3 and A1). S.V.P. and I.A.Yu. thank the Russian Foundation for Basic Research (Project No. 19-52-12046) and St. Petersburg State University (Grant No. 51125686). The research in Poland was partially supported by the Foundation for Polish Science through the IRA Programme, cofinanced by the EU within SG OP and by the National Science Centre through Grants No. 2017/25/B/ST3/02966 and 2018/30/M/ST3/00276.


\begin{thebibliography}{10}
	
	\bibitem{SpinBook} {\it Spin Physics in Semiconductors}, Springer International Publishing AG, Cham (2017), edited by M. I. Dyakonov.
	
	\bibitem{HakenWolfBook} H.~Haken and H.~C. Wolf, The Physics of Atoms and Quanta, Springer-Verlag, Berlin Heidelberg (1993).
	
	\bibitem{Semenov2003} Y.~G. Semenov and S.~M. Ryabchenko, Effects of photoluminescence polarization in semiconductor quantum wells subjected to an in-plane magnetic field, Phys. Rev. B \textbf{68}, 045322 (2003).
	
	\bibitem{KusrayevPRL1999} Y.~G. Kusrayev, A.~V. Koudinov, I.~G. Aksyanov, B.~P. Zakharchenya, T.~Wojtowicz, G.~Karczewski, and J.~Kossut, Extreme In-Plane Anisotropy of the Heavy-Hole $\mathit{g}$ Factor in (001)-CdTe/CdMnTe Quantum Wells, Phys. Rev. Lett. \textbf{82}, 3176 (1999).
	
	\bibitem{KoudinovPRB2004} A.~V. Koudinov, I.~A. Akimov, Y.~G. Kusrayev, and F.~Henneberger, Optical and magnetic anisotropies of the hole states in Stranski-Krastanov quantum dots, Phys. Rev. B \textbf{70}, 241305 (2004).
	
	\bibitem{Bogucki2016} A.~Bogucki, T.~Smole\'{n}ski, M.~Goryca, T.~Kazimierczuk, J.~Kobak, W.~Pacuski, P.~Wojnar, and P.~Kossacki, Anisotropy of in-plane hole $g$ factor in CdTe/ZnTe quantum dots, Phys. Rev. B \textbf{93}, 235410 (2016).
	
	\bibitem{Sirenko1997} A.~A. Sirenko, T.~Ruf, M.~Cardona, D.~R. Yakovlev, W.~Ossau, A.~Waag, and	G.~Landwehr, Electron and hole $g$ factors measured by spin-flip Raman scattering in CdTe/Cd${}_{1\ensuremath{-}x}$Mg${}_{x}$Te single quantum	wells, Phys. Rev. B \textbf{56}, 2114 (1997).
	
	\bibitem{Kusrayev2002} Y.~G. Kusrayev, A.~V. Koudinov, D.~Wolverson, and J.~Kossut, Anisotropy of spin-flip Raman scattering in CdTe/CdMnTe quantum wells, Phys. Status Solidi B \textbf{229}, 741 (2002).
	
	\bibitem{Koudinov2006} A.~V. Koudinov, N.~S. Averkiev, Y.~G. Kusrayev, B.~R. Namozov, B.~P. Zakharchenya, D.~Wolverson, J.~J. Davies, T.~Wojtowicz, G.~Karczewski, and J.~Kossut, Linear polarization of the photoluminescence of quantum wells subject to in-plane magnetic fields, Phys. Rev. B \textbf{74}, 195338 (2006).
	
	\bibitem{Debus2013} J.~Debus, D.~Dunker, V.~F. Sapega, D.~R. Yakovlev, G.~Karczewski, T.~Wojtowicz, J.~Kossut, and M.~Bayer, Spin-flip Raman scattering of the neutral and charged excitons confined in a CdTe/(Cd,Mg)Te quantum well, Phys. Rev. B \textbf{87}, 205316 (2013).
	
	\bibitem{Dahbashi2014} R.~Dahbashi, J.~H\"ubner, F.~Berski, K.~Pierz, and M.~Oestreich, Optical Spin Noise of a Single Hole Spin Localized in an (In,Ga)As Quantum Dot, Phys. Rev. Lett. \textbf{112}, 156601 (2014).
	
	\bibitem{Syperek2007} M.~Syperek, D.~R. Yakovlev, A.~Greilich, J.~Misiewicz, M.~Bayer, D.~Reuter, and A.~D. Wieck, Spin coherence of holes in GaAs/(Al,Ga)As quantum wells, Phys. Rev. Lett. \textbf{99}, 187401 (2007).
	
	\bibitem{ZhukovPSS2014} E.~A. Zhukov, D.~R. Yakovlev, A.~Schwan, O.~A. Yugov, A.~Waag, L.~W. Molenkamp, and M.~Bayer, Spin coherence of electrons and holes in ZnSe-based quantum wells studied by pump-probe Kerr rotation, Phys. Status Solidi B
	\textbf{251}, 1872 (2014).
	
	\bibitem{Gradl2014}	C.~Gradl, M.~Kempf, D.~Schuh, D.~Bougeard, R.~Winkler, C.~Sch\"uller, and T.~Korn, Hole-spin dynamics and hole $g$-factor anisotropy in coupled quantum well systems, Phys. Rev. B \textbf{90}, 165439 (2014).
	
	\bibitem{Belykh2016} V.~V. Belykh, D.~R. Yakovlev, J.~J. Schindler, E.~A. Zhukov, M.~A. Semina,	M.~Yacob, J.~P. Reithmaier, M.~Benyoucef, and M.~Bayer, Large anisotropy of electron and hole $g$ factors in infrared-emitting InAs/InAlGaAs self-assembled quantum dots, Phys. Rev. B \textbf{93}, 125302 (2016).
	
	\bibitem{AllenEberly} N.~Allen and J.~H. Eberly, {\it Optical Resonance and Two-Level Atoms}, Wiley, New York (1975).
	
	\bibitem{Lisin2012} V.~N. Lisin, A.~M. Shegeda, and K.~I. Gerasimov, Oscillations of the photon echo intensity in a pulsed magnetic field: Zeeman splitting in	LiLuF$_4$:Er$^{3+}$, JETP Letters \textbf{95}, 61 (2012).
	
	\bibitem{LangerPRL}	L.~Langer, S.~V. Poltavtsev, I.~A. Yugova, D.~R. Yakovlev, G.~Karczewski, T.~Wojtowicz, J.~Kossut, I.~A. Akimov, and M.~Bayer, Magnetic-field control of photon echo from the electron-trion system in a CdTe quantum well: shuffling coherence between optically accessible and inaccessible states, Phys. Rev. Lett. \textbf{109}, 157403 (2012).
	
	\bibitem{Invertor} S.~V. Poltavtsev, I.~A. Yugova, Y.~A. Babenko, I.~A. Akimov, D.~R. Yakovlev, G.~Karczewski, S.~Chusnutdinow, T.~Wojtowicz, and M.~Bayer, Quantum beats in the polarization of the spin-dependent photon echo from donor-bound excitons in CdTe/(Cd,Mg)Te quantum wells, arXiv:1911.08785 (2019).
	
	\bibitem{BirPikusBook} G.~L. Bir and G.~E. Pikus, {\it Symmetry and strain-induced effects in semiconductors}, Wiley, New York (1974).
	
	\bibitem{Marie1999} X.~Marie, T.~Amand, P.~Le~Jeune, M.~Paillard, P.~Renucci, L.~E. Golub, V.~D. Dymnikov, and E.~L. Ivchenko, Hole spin quantum beats in quantum-well structures, Phys. Rev. B \textbf{60}, 5811 (1999).
	
	\bibitem{Kiessling2006} T.~Kiessling, A.~V. Platonov, G.~V. Astakhov, T.~Slobodskyy, S.~Mahapatra, W.~Ossau, G.~Schmidt, K.~Brunner, and L.~W. Molenkamp, Anomalous in-plane magneto-optical anisotropy of self-assembled quantum dots, Phys.
	Rev. B \textbf{74}, 041301 (2006).
	
	\bibitem{Krizhanovskii2005} D.~N. Krizhanovskii, A.~Ebbens, A.~I. Tartakovskii, F.~Pulizzi, T.~Wright, M.~S. Skolnick, and M.~Hopkinson, Individual neutral and charged In$_x$Ga$_{1-x}$As-GaAs quantum dots with strong in-plane optical anisotropy, Phys. Rev. B \textbf{72}, 161312 (2005).
	
	
	\bibitem{SciRep2019} S.~V. Poltavtsev, Y.~V. Kapitonov, I.~A. Yugova, I.~A. Akimov, D.~R. Yakovlev, G.~Karczewski, M.~Wiater, T.~Wojtowicz, and M.~Bayer, Polarimetry of photon echo on charged and neutral excitons in semiconductor quantum wells, Sci. Rep. \textbf{9}, 5666 (2019).
	
	\bibitem{Poltavtsev2017} S.~V. Poltavtsev, M.~Reichelt, I.~A. Akimov, G.~Karczewski, M.~Wiater, T.~Wojtowicz, D.~R. Yakovlev, T.~Meier, and M.~Bayer, Damping of Rabi oscillations in intensity-dependent photon echoes from exciton complexes in a CdTe/(Cd,Mg)Te single quantum well, Phys. Rev. B \textbf{96}, 075306 (2017).
	
	\bibitem{Salewski2017} M.~Salewski, S.~V. Poltavtsev, I.~A. Yugova, G.~Karczewski, M.~Wiater, T.~Wojtowicz, D.~R. Yakovlev, I.~A. Akimov, T.~Meier, and M.~Bayer, High-resolution two-dimensional optical spectroscopy of electron spins, Phys. Rev. X \textbf{7}, 031030 (2017).
	
	\bibitem{PSS2018} S.~V. Poltavtsev, I.~A. Yugova, I.~A. Akimov, D.~R. Yakovlev, and M.~Bayer, Photon echo from localized excitons in semiconductor nanostructures, Phys. Solid State \textbf{60}, 1635 (2018).
	
	\bibitem{Saeed2018} F.~Saeed, M.~Kuhnert, I.~A. Akimov, V.~L. Korenev, G.~Karczewski, M.~Wiater, T.~Wojtowicz, A.~Ali, A.~S. Bhatti, D.~R. Yakovlev, and M.~Bayer, Single-beam optical measurement of spin dynamics in CdTe/(Cd,Mg)Te quantum wells, Phys. Rev. B \textbf{98}, 075308 (2018).
	
	\bibitem{Chamarro2019} G.~Garcia-Arellano, F.~Bernardot, G.~Karczewski, C.~Testelin, and M.~Chamarro, Spin relaxation time of donor-bound electrons in a CdTe quantum well, Phys. Rev. B \textbf{99}, 235301 (2019).
	
	\bibitem{Lawaetz1971} P.~Lawaetz, Valence-band parameters in cubic semiconductors, Phys. Rev. B \textbf{4}, 3460 (1971).
	
	\bibitem{Kiselev2001} A.~A. Kiselev, K.~W. Kim, and E.~Yablonovitch, In-plane light-hole $g$ factor in strained cubic heterostructures, Phys. Rev. B \textbf{64}, 125303
	(2001).
	
	\bibitem{Durnev2012} M.~V. Durnev, M.~M. Glazov, and E.~L. Ivchenko, Giant Zeeman splitting of light holes in GaAs/AlGaAs quantum wells, Phys. E \textbf{44}, 797 (2012).
	
	\bibitem{Glasberg1999} S.~Glasberg, H.~Shtrikman, I.~Bar-Joseph, and P.~C. Klipstein, Exciton exchange splitting in wide GaAs quantum wells, Phys. Rev. B \textbf{60}, R16295 (1999).
	
\end{thebibliography}

\end{document}